\documentclass[journal,draftcls,onecolumn,12pt,twoside]{IEEEtran}

\usepackage{amssymb, amsthm, enumerate, array, graphicx, changepage}
\usepackage[utf8]{inputenc} 
\usepackage[T1]{fontenc}
\usepackage{url}
\usepackage{ifthen}
\usepackage{cite}
\usepackage[cmex10]{amsmath} 
\usepackage{xcolor}
\usepackage{pgfplots}
\pgfplotsset{compat=1.18}
\usepackage{booktabs}
\usepackage{tabularx}
\usepackage{comment}
\usepackage{multirow}
\usepackage{makecell}
\usepackage{threeparttable}

\newtheorem{theorem}{Theorem}

\newtheorem{corollary}{Corollary}
\newtheorem{lemma}{Lemma}
\newtheorem{definition}{Definition}
\newtheorem{example}{Example}

\newtheorem*{definition*}{Definition}

\newtheoremstyle{customplain}
  {\topsep}   % 上方間距
  {\topsep}   % 下方間距
  {\itshape}  % 內文內建字體 (若想用正體可改為 \normalfont)
  {}          % 縮排量
  {\bfseries} % 標題字體 (粗體)
  {.}         % 標題後的標點符號
  {5pt plus 1pt minus 1pt} % 標題後的空白
  {#1 \thmnote{#3}} % 核心關鍵：將預設的 (#3) 改為 #3，直接移除括號！

\theoremstyle{customplain}
\newtheorem*{example*}{Example}

\newcommand{\CRT}{\mathrm{CRT}}
\newcommand{\C}{\mathcal{C}}
\newcommand{\Z}{\mathbb{Z}}
\newcommand{\E}{\mathbb{E}}
\newcommand{\F}{\mathbb{F}}
\newcommand{\I}{\mathcal{I}}
\newcommand{\A}{\mathcal{A}}

\newcommand{\s}{\mathbf{s}}

\newcommand{\ssum}{\mathsf{sum}}
\newcommand{\PP}{\text{P}}
\newcommand{\NP}{\text{NP}}

\begin{document}

\title{User-Irrepressible Sequences for Multiple-Packet Reception}

\author{
{Yen-Ling Shih, Yuan-Hsun Lo,~\IEEEmembership{IEEE Member}, \\ Yijin Zhang,~\IEEEmembership{IEEE Senior Member}, Ying Miao}
\thanks{
This work was supported in part by the National Science and Technology Council of Taiwan under Grant 114-2628-M-153-001-MY3, and in part by the Japan Society for the Promotion of Science Grant-in-Aid for Scientific Research (C) under Grant 24K06824.
}
\thanks{Y.-L. Shih and Y. Miao are with the Faculty of Engineering, Information and Systems, University of Tsukuba, Tsukuba (email: shih2862@gmail.com, miao@sk.tsukuba.ac.jp)}
\thanks{Y.-H. Lo is with the Department of Mathematical Sciences, National Chengchi University, Taipei (email: yhlo0830@gmail.com)}
\thanks{Y. Zhang is with the School of Electronic and Optical Engineering, Nanjing University of Science and Technology, Nanjing (email: yijin.zhang@gmail.com)}
}

\maketitle

%%%%%%%%%%%%%%%%%%%%%%%%%%%%%%%%%%%%%%%%%%%%%%%%
\begin{abstract}
This paper studies deterministic transmission schemes based on protocol sequences for slot-synchronous multiple-packet reception (MPR) channels without feedback or synchronization of sequence phases. 
Each user schedules its transmissions according to a preassigned periodic binary sequence. A sequence set has the user-irrepressible property with MPR capability $\gamma$ ($\gamma$-UI) if every user is guaranteed at least one successful transmission per period, regardless of the relative cyclic shifts. 
We propose a unified explicit construction based on the Chinese Remainder Theorem (CRT) for arbitrary MPR capability $\gamma\geq 2$. 
To the best of our knowledge, this is the first unified CRT-based construction framework for $\gamma$-UI sequences over this full range, without requiring conflict-avoiding codes with prescribed parameters. 
We establish the $\gamma$-UI property through bounds on pairwise Hamming cross-correlations and show that suitably selected subsets support one additional active user without increasing the sequence length. 
We also derive exact average individual delay expressions over all possible cyclic shifts and extend the delay analysis to $\gamma$-UI sequence sets obtained from conflict-avoiding codes.
\end{abstract}


%%%%%%%%%%%%%%%%%%%%%%%%%%%%%%
\section{Introduction}

%%%%%%%%%%%%%%%%%%%%%%%%%%%%%%
\subsection{Background}

In 1985, Massey and Mathys proposed the idea of using a set of deterministic and periodic $(0,1)$-binary sequences with a common period and prescribed Hamming cross-correlation properties, called \textit{protocol sequences}~\cite{MM85}, for multiple access channels without feedback. 
In a protocol-sequence-based scheme, each user reads out the entries from the assigned protocol sequence sequentially based on its local clock, sends a packet if the sequence value is $1$, and remains silent otherwise.
Although the channel is slot-synchronous, the users need not synchronize the starting phases of their assigned protocol sequences. 
Unlike probabilistic schemes, protocol sequences can offer the \textit{user-irrepressible (UI)} property~\cite{SWSC09,SZW10}: each assigned sequence guarantees its user at least one conflict-free transmission per period, regardless of the time offsets among users.
%The absence of sequence-phase synchronization and feedback requirements, together with the UI guarantee, makes protocol sequences particularly attractive for mission-critical communications in resource-constrained networks~\cite{LWLZC26}.
The demand for reliable and timely communication in IoT applications motivates access schemes with low coordination overhead~\cite{MXXPPV19,PAADH23}.
By avoiding sequence-phase synchronization and feedback while providing a UI guarantee, protocol sequences offer a promising approach to timely status updates in resource-constrained networks~\cite{LWLZCX23,LWLZC26}.

Most existing work on protocol sequences adopts the single-packet reception (SPR) channel, which assumes that a packet can
be correctly received if and only if no other packets are transmitted during its transmission~\cite{MM85,YK95,Wong07,SWSC09,SW10,SZW10,WSLWS13,WSWS14,ZSWS14,ZLSW19,LWZWX23}.
The SPR model, however, is becoming increasingly restrictive, as physical-layer advances endow receivers with multiple-packet reception (MPR) capability~\cite{NMT05,LE06}.
In this setting, UI sequences have been investigated in~\cite{ZLWS16,ZLSW14,ZLLWXZ24,SXWLZM26} under the widely used $\gamma$-MPR model, where a packet is successfully received if and only if at most $\gamma-1$ other packets are transmitted concurrently with it. 
The introduction of MPR brings new challenges to both sequence construction and performance analysis. 
From the construction perspective, collisions no longer need to be completely avoided; instead, the transmission patterns must be designed so that, regardless of the relative time offsets, every user has at least one transmission slot overlapped by fewer than $\gamma$ interfering users. 
Therefore, the construction problem shifts from controlling pairwise collisions to controlling the multiplicity of overlaps caused collectively by multiple users, making explicit constructions with short sequence lengths more challenging. 
From the performance-analysis perspective, MPR also complicates the characterization of throughput and access delay. 
The throughput of a user depends not only on the pairwise cross-correlations with other sequences, but also on how the overlaps caused by different interfering users accumulate at the same transmission positions. 
Access delay is even more sensitive to the sequence structure, since it depends on when the first successful transmission occurs and thus requires the joint success behavior of successive transmission opportunities to be characterized.

Existing approaches to constructing UI sequences for $\gamma$-MPR channels ($\gamma$-UI sequences) can be broadly classified into two categories.
The first approach relies on conflict-avoiding codes (CACs), which can be used to obtain $\gamma$-UI sequences for general MPR capability~\cite{ZLSW14,ZLLWXZ24}.
However, this approach requires CACs with suitable parameters, whose existence and explicit constructions are not fully understood in general.
The second approach constructs $\gamma$-UI sequences directly, without relying on the existence of a corresponding CAC. 
Explicit constructions of this type have primarily focused on $\gamma=1$ and $\gamma=2$.

For $\gamma=1$, several constructions have been developed, including shift-invariant sequences~\cite{MM85}, wobbling sequences~\cite{Wong07}, and CRT-based sequences~\cite{SWSC09}. 
For $\gamma=2$, the classical prime sequences~\cite{YK95} provide a construction for a prime number of users. 
More recently, our preliminary work~\cite{SXWLZM26} introduced new prime sequences and 2-CRT sequences for prime and composite parameters, respectively. 
Nevertheless, explicit constructions specifically designed to exploit the MPR capability for general $\gamma\geq 3$, without requiring a CAC as an intermediate construction, remain limited. 
This motivates the development of a unified construction framework for general MPR capabilities.

%%%%%%%%%%%%%%%%%%%%%%%%%%%%%%
\subsection{Preliminaries and Related Work}

For any positive integer $L$, let $\Z_L\triangleq\{0,1,\ldots,L-1\}$ denote the ring of residues modulo $L$.
Let $\s=s(0)s(1)\cdots s(L-1)$ be a binary sequence of length $L$.
The number of $1$'s in $\s$ is called the \emph{Hamming weight} (\emph{weight} for short) of $\s$, denoted by 
$w(\s)$.
For any $\tau\in\Z_L$, let $\s^{\tau}=s^\tau(0)s^\tau(1)\cdots s^\tau(L-1)$ denote the cyclic shift of $\s$ by $\tau$ positions, defined by $s^\tau(t)=s(t-\tau)$, where the indices are taken modulo $L$.
%For any non-negative integer $\tau$, denote by $\s^{\tau}\triangleq s(\tau)s(\tau+1)\cdots s(\tau+L-1)$ the \emph{cyclic shift} of $\s$ by $\tau$, where the addition is performed in $\Z_L$.
%For convenience, $\tau$ is normalized as in $\Z_L$.
Since $\s$ can be represented by its \emph{characteristic set} 
\begin{align*}
\I_{\s}\triangleq\{t\in\Z_L:\,s(t)=1\},
\end{align*}
the characteristic set of $\s^{\tau}$ is given by $\I_{\s}+\tau\triangleq\{t+\tau\bmod L:\, t\in\I_{\s}\}$.
Note that $|\I_{\s}+\tau|=w(\s)$ for any $\tau\in\Z_L$.
Furthermore, define the \emph{set of (non-zero) differences} of $\s$ by $d^*(\s)\triangleq\{x-y\bmod L:\,x,y\in\I_{\s}, x\neq y\}$.

For any two distinct sequences $\s_i,\s_j$ and any $\tau\in\Z_L$, define $H_{\s_i,\s_j}(\tau)\triangleq |\I_{\s_i}\cap(\I_{\s_j}+\tau)|$.
The \emph{Hamming cross-correlation} of $\s_i$ and $\s_j$ is defined by 
\begin{align*}
H_{\s_i,\s_j}\triangleq\max_{\tau\in\Z_L}H_{\s_i,\s_j}(\tau).
\end{align*}
By definition, $H_{\s_i,\s_j}=H_{\s_j,\s_i}$.
Obviously, $H_{\s_i,\s_j}=1$ if and only if $d^*(\s_i)\cap d^*(\s_j)=\emptyset$.
%Obviously, $H_{\s_i,\s_j}=1$ (resp., $H_{\s_i,\s_j}=2$) if and only if $d^*(\s_i)\cap d^*(\s_j)=\emptyset$ (resp., $|d^*(\s_i)\cap d^*(\s_j)|=1$).

For any set $\A$ and a positive integer $m$, let $\A^{m}$ denote the multiset containing $m$ copies of all elements in $\A$.
The formal definition of UI sequences under $\gamma$-MPR, called $\gamma$-UI sequences, is given below. 

\begin{definition}\label{def:UI-MPR}\rm
Let $\C=\{\s_1,\ldots,\s_N\}$ be a set of binary sequences with common length $L$.
The set $\C$ is said to have the $\gamma$-UI property if for any $j$, one has
\begin{equation}\label{eq:UI-MPR-def}
\bigcup_{k\neq j}\left(\I_{\s_k} +\tau_k\right) \nsupseteq (\I_{\s_j})^\gamma,
\end{equation}
for any $\boldsymbol{\tau}=(\tau_1,\ldots,\tau_N)\in\Z_L^N$, where the characteristic sets are considered as multisets.
\end{definition}

An approach to constructing $\gamma$-UI sequences for general MPR capability is based on \emph{conflict-avoiding codes (CACs)}. 
A set of binary sequences $\mathcal{C}=\{\s_1,\ldots,\s_N\}$ with common length $L$ is called a CAC of weight $w$ if $w(\s_i)=w$ for all $i$ and $H_{\s_i,\s_j}=1$ for all $i\neq j$. 
As shown in \cite{ZLSW14}, the Pigeonhole Principle implies that $\mathcal{C}$ has the $\gamma$-UI property whenever $N\leq\gamma w$. 
Therefore, to construct a set of $N$ $\gamma$-UI sequences, it suffices to find a CAC with $N$ codewords and weight $w=\lceil N/\gamma\rceil$. 
However, the existence of CACs with prescribed length, weight, and code size is not fully characterized, and explicit constructions with short sequence lengths are known only for certain parameter regimes. 
For further results on CACs, we refer the reader to~\cite{LT05,JMJTT07,MMSJ07,MFU09,SWC10,SW10CAC,FLM10,FLS14,LWXZ25}.

We next review existing direct constructions of $\gamma$-UI sequences that do not rely on CACs. 
These constructions have mainly been developed for $\gamma=1$ and $\gamma=2$.

For the case of $\gamma = 1$, various construction frameworks for $1$-UI sequences have been well-established \cite{MM85,YK95,Wong07,SW10,SZW10}. 
Notable sequence families include shift-invariant sequences~\cite{MM85}, extended prime sequences~\cite{YK95}, wobbling sequences~\cite{Wong07}, CRT sequences~\cite{SWSC09,SW10}, and $\CRT_p$ sequences~\cite{SZW10}.
In particular, CRT sequences and $\CRT_p$ sequences are constructed based on the \textit{Chinese Remainder Theorem (CRT)} and currently offer the shortest sequence lengths among all known $1$-UI sequences.
For $\gamma = 2$, the available constructions depend on the construction parameter. 
For a prime parameter $p$, classical prime sequences~\cite{YK95} and new prime sequences~\cite{SXWLZM26} provide 2-UI constructions. 
For a composite parameter $K$, our preliminary work~\cite{SXWLZM26} introduced 2-CRT sequences and established the 2-UI property for any $K$ sequences chosen from the constructed family. 
In this paper, Theorem~\ref{thm:gamma-CRT-UI} further shows that any $K+1$ sequences from this family have the 2-UI property provided that the selected subset contains at least one of the two distinguished sequences $\s_0$ and $\s_{p_K}$, defined in Section~\ref{sec:construction}. 
Thus, one additional active user can be supported under this selection condition without increasing the sequence length. 
Table~\ref{tab:known_constructions} summarizes the known direct constructions and the improved user guarantees established in this paper.
Note that $p_K$ refers to the smallest prime number greater than or equal to $K$.
According to the Bertrand-Chebyshev Theorem, $K \leq p_K < 2K$.
The upper bound of $p_K$ was improved in~\cite{Nagura52} to be $6K/5$ when $K\geq 25$.

\begin{table}[h]
\centering
\begin{threeparttable}
\caption{Summary of Known Direct Constructions for $\gamma$-UI Sequences} 
\label{tab:known_constructions}
\begin{tabular}{c|lcllcc}
\textbf{$\gamma$} & \textbf{Construction} & \textbf{Reference} & \textbf{Parameter} &  \makecell[l]{\textbf{Sequence}\\[-3pt]\textbf{Length}} &  \makecell[l]{\textbf{Number of}\\[-3pt]\textbf{Constructed}\\[-3pt]\textbf{Sequences}} &  \makecell[l]{\textbf{Number of}\\[-3pt]\textbf{Applicable}\\[-3pt]\textbf{Users} $N$} \\ \hline

\multirow{5}{*}{$1$}
& Shift-invariant Sequences & \cite{MM85} & Arbitrary $K$  & $2^{K}$ & $K$ & $K$ \\
& Wobbling Sequences & \cite{Wong07} & Prime $p$ & $p^4$ & $p$ & $p$ \\
& Extended Prime Sequences & \cite{YK95} & Prime $p$ & $p(2p-1)$ & $p$ & $p$ \\
& CRT Sequences & \cite{SWSC09,SW10} & Arbitrary $K$ & $p_K(2K-1)$ & $p_K+1$ & $K$ \\
& $\CRT_p$ Sequences & \cite{SZW10} & Prime $p$ & $2p(p-1)$ & $p+1$ & $p$ \\ \hline

\multirow{3}{*}{$2$}
& Prime Sequences & \cite{YK95} & Prime $p$ & $p^2$ & $p$ & $p$ \\ 
& {New Prime Sequences} & \cite{SXWLZM26} & Prime $p$ & $p^2$ & $p+1$ & $p$ \\ 
& & This work (App.~\ref{app:2UI_proof}) & Prime $p$ & $p^2$ & $p+1$ & $p+1$ \\
& 2-CRT sequences & \cite{SXWLZM26} & Composite $K$ & $p_KK$ & $p_K+1$ & $K$ \\ 
& & This work (Thm.~\ref{thm:gamma-CRT-UI}) & Composite $K$ & $p_KK$ & $p_K+1$ & $K+1$\tnote{a} \\ \hline
\end{tabular}
\begin{tablenotes}[flushleft]
\footnotesize
\item[a] The guarantee for $N=K+1$ requires the selected subset to contain at least one of $\s_0$ and $\s_{p_K}$; any subset of at most $K$ sequences satisfies the $2$-UI property.
\end{tablenotes}
\end{threeparttable}
\end{table}

It should be noted that, in applications, the number of constructed sequences and the number of applicable users can be viewed as the number of supported \emph{potential users} and the number of \emph{active users}, respectively~\cite{LT05}.

For completeness, Appendix~\ref{app:2UI_proof} strengthens the result for the new prime sequences in~\cite{SXWLZM26} by establishing the 2-UI property for the full family of $p+1$ sequences. 
Appendix~\ref{app:isomorphism} shows that a specified subset of $p$ sequences coincides with the classical prime sequences in~\cite{YK95} after reindexing, thereby providing an alternative proof of their 2-UI property.

%%%%%%%%%%%%%%%%%%%%%%%%%%%%%%
\subsection{Contributions}

As discussed above, existing approaches to constructing $\gamma$-UI sequences for general MPR capability often rely on CACs, whose existence and construction for prescribed parameters remain challenging. 
In this paper, we extend the CRT methodology in~\cite{SXWLZM26} to develop a direct and explicit construction of $\gamma$-UI sequences for general MPR capabilities. 
The construction applies to all integers $K\geq\gamma\geq 2$, with the additional requirement that $K$ be composite when $\gamma=2$.
%Our construction applies to arbitrary $\gamma\geq 3$ and arbitrary system parameter $K$, as well as to $\gamma=2$ when $K$ is composite. 
Unlike CAC-based approaches, the proposed framework does not require the existence of a CAC with prescribed parameters.
To the best of our knowledge, this is the first unified CRT-based construction framework tailored to arbitrary MPR capabilities.

The main contributions of this paper are summarized as follows:
\begin{itemize}
\item \textbf{General Construction and Cross-Correlation Bounds:} 
We propose a direct CRT-based construction of $\gamma$-UI sequence sets for general MPR capabilities. 
Through a structural analysis of the constructed sequences, we derive upper bounds on their pairwise Hamming cross-correlations.

\item \textbf{$\gamma$-UI Property and Extended User Support:} 
The proposed construction contains $p_K+1$ sequences, and we prove that every subset of at most $K$ sequences has the $\gamma$-UI property. 
We further establish this property for every subset of $K+1$ sequences containing at least one of the two distinguished sequences $\s_0$ and $\s_{p_K}$. 
Thus, the construction supports up to $K+1$ active users under the stated selection condition. 
For $\gamma=2$ and composite $K$, this extends the guarantee in~\cite{SXWLZM26} without increasing the sequence length.

\item \textbf{Exact Average Individual Delay Analysis:} 
Under independent uniform cyclic shifts, we derive exact expressions for the average individual delay of the proposed sequences. 
We obtain a simplified expression for $\s_0$ and develop a collision-counting method for the remaining sequences. 
We further extend the analysis to $\gamma$-UI sequence sets obtained from constant-weight CACs.

\item \textbf{Additional Results for Prime Sequences:} 
We establish the $2$-UI property of the full family of $p+1$ new prime sequences and prove that a specified subset of $p$ sequences coincides with the classical prime sequences after reindexing.
\end{itemize}

The remainder of this paper is organized as follows. 
Section~\ref{sec:construction} presents the $\gamma$-CRT construction. 
Section~\ref{sec:gamma-UI} establishes the cross-correlation bounds and the $\gamma$-UI property. 
Section~\ref{sec:delay-0} derives the average individual delay of $\s_0$ and extends the analysis to CAC-based sequences, while Section~\ref{sec:delay-g} treats the remaining $\gamma$-CRT sequences. 
Section~\ref{sec:conclusion} concludes the paper and discusses future research directions. 
Appendices~\ref{app:2UI_proof} and \ref{app:isomorphism} establish the additional results for new prime and classical prime sequences.

%%%%%%%%%%%%%%%%%%%%%%%%%%%%%%
\section{Construction of $\gamma$-CRT Sequences}\label{sec:construction}

We use $\gamma$-CRT$(K)$ to denote the set of sequences constructed for given integers $K\geq \gamma\geq 2$ in this section.
The common sequence length is $L=p_K\lceil 2K/\gamma\rceil$.
The set $\gamma$-CRT$(K)$ consists of $p_K+1$ sequences $\s_0,\s_1,\ldots,\s_{p_K}$, which can be assigned to $p_K+1$ potential users, whose characteristic sets are given below.

\begin{definition}\label{def:gamma-CRT} \rm
Let $K\geq\gamma\geq 2$ be integers, with $K$ composite when $\gamma=2$.
For $g=0,1,...,p_K-1$, let
\begin{equation}\label{eq:gamma-CRT-def-g}
\widehat{\I}_{\s_g}\triangleq\{j(g,1) \in \Z_{p_K}\times\Z_{\lceil 2K/\gamma\rceil}:j=0,1,...,K-1 \};
\end{equation} 
and let
\begin{equation}\label{eq:gamma-CRT-def-p}
\widehat{\I}_{\s_{p_K}}\triangleq\{j(1,0) \in \Z_{p_K}\times\Z_{\lceil 2K/\gamma\rceil}:j=0,1,...,K-1 \}.
\end{equation}
For $0\leq g\leq p_K$, the characteristic set of $\s_g$ is given by
\begin{equation}\label{eq:gamma-CRT-def}
\I_{\s_g} := \{t\in\Z_{p_K\cdot\lceil 2K/\gamma\rceil} :\, (t \bmod{p_K}, t \bmod{\lceil 2K/\gamma\rceil})\in\widehat{\I}_{\s_g}\}.
\end{equation}
The elements $(g,1)$ in \eqref{eq:gamma-CRT-def-g} and $(1,0)$ in~\eqref{eq:gamma-CRT-def-p} are called \emph{generators}.
\end{definition}

Note that the definition in~\eqref{eq:gamma-CRT-def} is well-defined since $\Z_{p_K\cdot\lceil 2K/\gamma\rceil}\cong\Z_{p_K}\times\Z_{\lceil 2K/\gamma\rceil}$ due to $\gcd(p_K,\lceil 2K/\gamma\rceil)=1$, with the assumption that $\gamma\geq 2$ and $K$ non-prime if $\gamma=2$.
More precisely, the mapping $\theta:\Z_{p_K\cdot\lceil 2K/\gamma\rceil}\to\Z_{p_K}\times\Z_{\lceil 2K/\gamma\rceil}$ defined by
\begin{equation}\label{eq:CRT}
\theta(t) = (t\bmod p_K, t\bmod \lceil 2K/\gamma\rceil)
\end{equation}
is a bijection.
This correspondence is known as the \textit{Chinese Remainder Theorem}.
In this fashion, the cyclic shift $\I_{\s_g}+\tau$ corresponces to 
\begin{align*}
\widehat{\I}_{\s_g}+(\tau_x,\tau_y),
\end{align*}
where $\tau\equiv\tau_x\bmod p_K$ and $\tau\equiv\tau_y\bmod \lceil 2K/\gamma\rceil$.
Therefore, we also call $\widehat{\I}_{\s_g}$ the characteristic set of $\s_g$ with generator $\theta^{-1}(\mathfrak{g})$, where $\mathfrak{g}$ is the generator of $\widehat{\I}_{\s_g}$.

In $\gamma$-CRT$(K)$, we have $w(\s_0)=\lceil 2K/\gamma\rceil$ and $w(\s_g)=K$ for $1\leq g\leq p_K$.
The characteristic set of the special case $\s_0$ is
\begin{equation}\label{eq:gamma-CRT-def-0_chara}
\I_{\s_{0}} = \{0, p_K, 2p_K, \ldots, (\lceil 2K/\gamma\rceil-1)p_K\},
\end{equation} 
as
\begin{equation}\label{eq:gamma-CRT-def-0}
\widehat{\I}_{\s_{0}} = \{(0,0), (0,1), (0,2), \ldots, (0,\lceil 2K/\gamma\rceil-1)\}.
\end{equation} 
Note that for the case that $\gamma=2$ and $K$ is composite, $\gamma$-CRT$(K)$ recovers the $2$-CRT$(K)$ construction in~\cite{SXWLZM26}, thereby offering a natural extension to arbitrary $\gamma\geq 2$.

\begin{example}\label{ex:gamma=3}\rm
Consider $\gamma=3$ and $K=8$.
In this case, we have $p_K=11, \lceil 2K/\gamma\rceil = 6$, and $L=66$.
The generators of the $p_K+1=12$ sequences are $(0,1),(1,1),\ldots,(10,1)$, and $(1,0)$ in $\Z_{11}\times\Z_6$, which correspond to $55, 1, 13, 25, 37, 49, 61, 7, 19, 31, 43,$ and $12$ in $\mathbb{Z}_{66}$ under the CRT correspondence~\eqref{eq:CRT}.
The characteristic sets of the constructed $12$ sequences are:
\begin{align*}
\I_{\s_0} &= \{0,11,22,33,44,\mathbf{55}\}, & \I_{\s_1} &= \{0,\mathbf{1},2,3,4,5,6,7\}, \\
\I_{\s_2} &= \{0,12,\mathbf{13},25,26,39,52,65\}, & \I_{\s_3} &= \{0,9,18,\mathbf{25},34,43,50,59\},\\
\I_{\s_4} &= \{0,8,16,24,\mathbf{37},45,53,61\}, & \I_{\s_5} &= \{0,13,15,30,32,47,\mathbf{49},64\},\\
\I_{\s_6} &= \{0,31,36,41,46,51,56,\mathbf{61}\}, & \I_{\s_7} &= \{0,\mathbf{7},14,21,28,35,42,49\},\\
\I_{\s_8} &= \{0,1,10,\mathbf{19},29,38,48,57\}, & \I_{\s_9} &= \{0,19,23,27,\mathbf{31},54,58,62\},\\
\I_{\s_{10}} &= \{0,17,20,37,40,\mathbf{43},60,63\}, & \I_{\s_{11}} &= \{0,6,\mathbf{12},18,24,36,48,60\}. &
\end{align*}
The generators are highlighted in bold.
Note that there are only $6$ elements in $\I_{\s_0}$.
\end{example}

%%%%%%%%%%%%%%%%%%%%%%%%%%%%%%
\section{Cross-Correlation Bounds and the $\gamma$-UI Property}\label{sec:gamma-UI}

We now establish sufficient conditions for subsets of $\gamma$-CRT$(K)$ to have the $\gamma$-UI property. 
We first derive pairwise cross-correlation bounds and then use these bounds to control the total interference at the transmission positions of each user.
%To establish the $\gamma$-UI property for the proposed construction, we first derive upper bounds on the Hamming cross-correlation between any two distinct sequences in $\gamma$-CRT$(K)$.

\begin{lemma}\label{lem:gamma-UI}
Consider the sequence set $\gamma$-CRT$(K)$ defined in Definition~\ref{def:gamma-CRT}.
For any two distinct indices $g$ and $h$, one has
\begin{enumerate}[(i)]
\item $H_{\s_g, \s_h}=1$ if $g=0$ or $h=0$;
\item $H_{\s_g, \s_h} \leq \lceil\gamma/2\rceil$ if $g,h\neq 0$ and exactly one of them is $p_K$; and
\item $H_{\s_g, \s_h} \leq \gamma$ if $0 < g,h < p_K$.
\end{enumerate}
\end{lemma}
\begin{proof}
For the sake of convenience, let $q = \lceil 2K/\gamma \rceil$.
Assume $1\leq g\leq p_K$.
By~\eqref{eq:gamma-CRT-def-g} and \eqref{eq:gamma-CRT-def-p}, the first components of the ordered pairs in $\widehat{\I}_{\s_g}$ form a subset of $\Z_{p_K}$ consisting of $K$ distinct elements.
Adding any integer $\tau$ to these components modulo $p_K$ preserves this property, yielding another subset of $\Z_{p_K}$ with the same distinctness.
Therefore, by~\eqref{eq:gamma-CRT-def-0}, one has $H_{\s_g, \s_0}=1$.
This completes the proof of (i).

Then, we consider two sequences $\s_g$ and $\s_h$ for $1\leq  g, h\leq p_K-1$ with $g\neq h$.
Suppose that the $m$-th packet of $\s_g$ and the $n$-th packet of $\s_h$ collide under a relative time shift $\tau = (\tau_x, \tau_y) \in \mathbb{Z}_{p_K} \times \mathbb{Z}_q$. 
By~\eqref{eq:gamma-CRT-def-g}, this collision leads to the following system of congruences:
\begin{align}
    m \cdot g &\equiv n \cdot h + \tau_x \pmod{p_K}, \label{eq:gamma_x} \\
    m &\equiv n + \tau_y \pmod q, \label{eq:gamma_y}
\end{align}
where $m, n \in \{0, 1, \dots, K-1\}$. 
From \eqref{eq:gamma_y}, the difference between both sides is a multiple of $q$. Thus, there exists an integer $c$ such that:
\begin{equation}\label{eq:sub_c}
    n = m - \tau_y + c \cdot q.
\end{equation}
Substituting \eqref{eq:sub_c} into \eqref{eq:gamma_x} yields
\begin{align*}
    m \cdot g \equiv (m - \tau_y + c \cdot q)h + \tau_x \pmod{p_K},
\end{align*}
which implies that
\begin{align*}
    m(g - h) \equiv \tau_x - \tau_y h + c \cdot q \cdot h \pmod{p_K}.
\end{align*}
Since $p_K$ is a prime and $g \neq h$, the term $(g-h)$ is invertible in the finite field $\mathbb{F}_{p_K}$. Multiplying both sides by $(g-h)^{-1}$ isolates $m$, namely,
\begin{equation}\label{eq:m_unique}
    m \equiv (g - h)^{-1} (\tau_x - \tau_y h + c \cdot q \cdot h) \pmod{p_K}.
\end{equation}
For a fixed time shift $(\tau_x, \tau_y)$, the right-hand side of \eqref{eq:m_unique} is a constant determined entirely by $c$. 
This indicates that each admissible value of $c$ corresponds to at most one overlapping transmission position, with index $m$. 
Consequently, the maximum number of collided packets between $\s_g$ and $\s_h$ is upper-bounded by the number of possible values for $c$.

To find the range of $c$, we rewrite \eqref{eq:sub_c} as $c \cdot q = n - m + \tau_y$. Since $m, n \in \{0, 1, \dots, K-1\}$, the difference $(n - m)$ is bounded by $-(K-1) \le n - m \le K-1$. 
Thus, the value $c \cdot q$ must fall within the interval $[-(K-1)+\tau_y,(K-1)+\tau_y]$, which is of length $2K-2$.
The maximum number of multiples of $q$ (i.e., possible values of $c$) within a closed interval of length $2K-2$ is given by $\lfloor (2K-2)/q \rfloor + 1$. 

By definition, $q = \lceil 2K/\gamma \rceil \ge 2K/\gamma$, which implies $2K/q \le \gamma$. We then have:
\begin{equation*}
\frac{2K - 2}{q} = \frac{2K}{q} - \frac{2}{q} \le \gamma - \frac{2}{q}.
\end{equation*}
Since $\gamma - 2/q < \gamma$ and $\gamma$ is an integer, we obtain $\lfloor (2K-2)/q \rfloor \le \gamma - 1$. 
Therefore, the number of possible values for $c$ is at most $(\gamma - 1) + 1 = \gamma$. 
This proves $H_{s_g, s_h} \le \gamma$, confirming the third case (iii).

Finally, consider the special case when $1\leq g\leq p_K-1$ and $h=p_K$.
Since $\widehat{\mathcal{I}}_{\s_{p_K}}$ is generated by $(1,0)$, the congruence~\eqref{eq:gamma_y} is simplified to 
\begin{align*}
    m \equiv \tau_y \pmod q,
\end{align*}
which implies that $m=c\dot q+\tau_y$ for some integer $c$.
As $m\in\{0,1,\ldots,K-1\}$, given a fixed $\tau_y$, the number of possible values of $c$ (i.e., the number of collided packets between $\s_g$ and $\s_{p_K}$) is at most
\begin{align*}
    \left\lceil\frac{K}{q}\right\rceil = \left\lceil\frac{K}{\lceil 2K/\gamma \rceil} \right\rceil \leq \left\lceil\frac{\gamma}{2}\right\rceil,
\end{align*}
as desired.
\end{proof}

%We remark that the upper bounds of $H_{\s_g,\s_h}$ in Lemma~\ref{lem:gamma-UI}(ii)--(iii) are indeed achievable.
%In fact, shift pair $(\tau_x,\tau_y)$ attaining maximum cross-correlation can be explicitly characterized: one first identifies a candidate shift $\tau_y$ such that the interval $[-(K-1)+\tau_y,(K-1)+\tau_y]$ contains $\gamma$ (or $\gamma/2$ when considering $\s_{p_K}$) solutions for $c$, followed by solving $\tau_x$ via~\eqref{eq:gamma_x}--\eqref{eq:gamma_y}.

\begin{example*}[\ref{ex:gamma=3} (cont.)] \rm
Observe that $d^*(\s_0)=\{11k:\,k=1,2,3,4,5\}$ and $11k\notin d^*(\s_g)$ for $1\leq g\leq 11$.
It follows that $H_{\s_0,\s_g}=1$ for $1\leq g\leq 11$.
Moreover, it is not hard to verify that $H_{\s_{11},\s_g}\leq 2$ for $1\leq g\leq 10$ and $H_{\s_g,\s_h}\leq 3$ for $1\leq g\neq h\leq 10$.
\end{example*}

Based on the maximum cross-correlation derived above, we can verify the $\gamma$-UI property of the constructed set of sequences $\gamma$-CRT$(K)$.

\begin{theorem}\label{thm:gamma-CRT-UI}
Consider the sequence set $\gamma$-CRT$(K)$ defined in Definition~\ref{def:gamma-CRT}, and let $\C\subseteq \gamma\text{-CRT}(K)$. 
Then $\C$ has the $\gamma$-UI property if either of the following conditions holds.
\begin{enumerate}[(i)]
\item $|\C|\leq K$.
\item $|\C|=K+1$ and $\C\cap\{\s_0,\s_{p_K}\}\neq\emptyset$.
\end{enumerate}
\end{theorem}

\begin{proof}
For $\s_g\in\C$, the maximum number of overlaps from the remaining interfering sequences in $\mathcal{C}$ is bounded above by $\sum_{\s_h\in \C \setminus \{\s_g\}} H_{\s_g,\s_h}$
for any combination of time shifts, counting multiplicity.
According to ~\eqref{eq:UI-MPR-def}, by the Pigeonhole Principle, $\C$ has the $\gamma$-UI property if 
\begin{equation}\label{eq:gamma-UI-proof}
    \gamma \cdot |\widehat{\mathcal{I}}_{\s_g}| > \sum_{\s_h\in\C \setminus \{\s_g\}}  H_{\s_g,\s_h}, 
\end{equation}
for all $\s_g\in\C$.
To show that, we partition the elements in $\C$ into three classes: $\s_0, \s_{p_K}$, and $\s_g$ ($1\leq g<p_K$), and prove each case separately.
Note that $K\geq|\C|-1$ in either case (i) or case~(ii).

When $\s_0\in\C$, by Lemma~\ref{lem:gamma-UI}(i), the right-hand-side of~\eqref{eq:gamma-UI-proof} is $|\C|-1$.
Since $|\widehat{\mathcal{I}}_{\s_0}|=\lceil 2K/\gamma \rceil$, we have $\gamma\lceil 2K/\gamma \rceil\geq 2K>K\geq |\C|-1$, which satisfies~\eqref{eq:gamma-UI-proof}.

When $\s_{p_K}\in\C$, by Lemma~\ref{lem:gamma-UI}(ii), the right-hand-side of~\eqref{eq:gamma-UI-proof} is at most $\lceil\gamma/2\rceil(|\C|-1)$.
More precisely, it is bounded above by $\lceil\gamma/2\rceil(|\C|-1)$ if $\s_0\notin\C$ and $\lceil\gamma/2\rceil(|\C|-2)+1$ otherwise.
As $|\widehat{\mathcal{I}}_{\s_{p_K}}|=K$, we have $\gamma K \geq \gamma(|\C|-1) > \lceil\gamma/2\rceil(|\C|-1)$, provided $\gamma\geq 2$.

Consider $\s_g\in\C$ with $1\leq g<p_K$.
By Lemma~\ref{lem:gamma-UI}, the right-hand-side of~\eqref{eq:gamma-UI-proof} is at most $\gamma(|\C|-1)$.
Since $|\widehat{\mathcal{I}}_{\s_{g}}|=K$, we have $\gamma K > \gamma(K-1)\geq \gamma(|\C|-1)$. 
So, \eqref{eq:gamma-UI-proof} holds for case~(i), namely, $K\geq|\C|$.
As for case (ii), the right-hand-side of~\eqref{eq:gamma-UI-proof} is 
\begin{align*}
\sum_{\s_h\in \C \setminus \{\s_g\}} H_{\s_g,\s_h} \leq 
\begin{cases}
    \gamma(K-1) + 1, & \text{if } \s_0\in\C \text{ and } \s_{p_K}\notin\C, \\
    \gamma(K-1) + \lceil\gamma/2\rceil, & \text{if } \s_0\notin\C \text{ and } \s_{p_K}\in\C, \\
    \gamma(K-2) + 1 + \lceil\gamma/2\rceil, & \text{if } \s_0,\s_{p_K}\in\C.
\end{cases}
\end{align*}
Clearly, the inequality~\eqref{eq:gamma-UI-proof} holds for either case.
This completes the proof.
\end{proof}

Theorem~\ref{thm:gamma-CRT-UI} means that the ${\gamma}\text{-CRT}(K)$ defined in Definition~\ref{def:gamma-CRT} can support a system of up to $p_K+1$ potential users, where any group of up to $K$ users can be active.
Furthermore, if one user in this group is assigned the sequence $\s_0$ or $\s_{p_K}$, then the number of active users can even be $K+1$.

%%%%%%%%%%%%%%%%%%%%%%%%%%%%%%
\section{Average Individual Delay of $\s_0$ and CAC-Based Sequences}\label{sec:delay-0}

In this and the next section, we analyze average \emph{individual delay} under independent uniform cyclic shifts. 
Unless stated otherwise, let $\C\subseteq \gamma\text{-CRT}(K)$ be a $\gamma$-UI sequence set that satisfies one of the conditions in Theorem~\ref{thm:gamma-CRT-UI}.
Let $X_{\gamma;\C,g}$ denote the number of slots from slot $0$ to the first successful transmission of the user assigned $\s_g$. 
In particular, a successful transmission in slot $0$ has zero delay. 
In this section, we first analyze $\s_0$ and then extend the method to CAC-based sequence sets.

%In this section, we analyze the \textit{individual delay} of the $\gamma$-CRT scheme, defined as the number of slots from slot $0$ to the first successful transmission. 
%In particular, a successful transmission in slot $0$ has zero delay. 
%We first consider the distinguished sequence $\s_0$, whose transmission opportunities are equally spaced, and then analyze the remaining sequences.

%Let $\C\subseteq \gamma\text{-CRT}(K)$ be a $\gamma$-UI sequence set that satisfies the condition (i) or (ii) given in Theorem~\ref{thm:gamma-CRT-UI}.
%Let random variable $X_{\gamma;\C,g}$ denote the individual delay of $\s_g$ in $\C$ under the $\gamma$-MPR model.
To characterize the average individual delay, we assume that the cyclic shift of each sequence in $\C$ is chosen independently and uniformly at random from $\{0,1,\ldots,L-1\}$. 
%Thus, the cyclic shifts of the sequences are independent and identically distributed. 
The average individual delay of $\s_g$ in $\C$ under the $\gamma$-MPR model is then defined as
\begin{align*}
\E\!\left[X_{\gamma;\mathcal{C},g}\right],
\end{align*}
where the expectation is taken over all possible combinations of cyclic shifts of the sequences in $\C$.

%%%%%%%%%%%%%%%%%%%%%%%%%%%%%%
\subsection{Success Probabilities}\label{sec:delay-0-Pn}

Assume that $\s_0$ is in $\C$.
Let $q=\lceil 2K/\gamma\rceil$, so that $L=p_Kq$ and $\I_{s_0}=\{0,p_K,\ldots,(q-1)p_K\}$.

Fix a cyclic shift $\tau\in\Z_L$ of $\s_0$, and consider the shifted characteristic set ${\I}_{\s_0}+\tau = \{x_1(\tau),x_2(\tau),\ldots,x_q(\tau)\}$, where $0\leq x_1(\tau)<x_2(\tau)<\cdots<x_q(\tau) <L$.
For $1 \le n \le q$, let $P_n$ denote the probability that the user equipped with $\s_0$ transmits at least one successful packet at time slots $x_1(\tau),x_2(\tau),\ldots,x_n(\tau)$.
The explicit formula for $P_n$ is given in Lemma~\ref{lem:P_n_0}, which will also indicate that the probability is independent of $\tau$.

Consider an $m$-tuple $\boldsymbol{\sigma}=(\sigma_1,\ldots,\sigma_m)$ whose entries are from the set $\{0,1,\ldots,\gamma-1\}$. 
Let $\ssum(\boldsymbol{\sigma)}\triangleq\sum_{i=1}^{m}\sigma_i$ denote the sum of all entries in $\boldsymbol{\sigma}$. 
We further use $\binom{M}{\boldsymbol{\sigma}}$ to denote the multinomial coefficient $\binom{M}{\sigma_1}\binom{M-\sigma_1}{\sigma_2}\cdots\binom{M-(\sigma_1+\cdots\sigma_{m-1})}{\sigma_m}$, where $M$ is a given integer that is not less than $\ssum(\boldsymbol{\sigma})$.

\begin{lemma}\label{lem:P_n_0}
For $n > (|\C|-1)/\gamma$, one has $P_n=1$, and 
\begin{equation}\label{eq:P_n_0}
P_n = \sum_{m=1}^{n}(-1)^{m-1}\binom{n}{m} \sum_{\boldsymbol{\sigma}\in\{0,\ldots,\gamma-1\}^m}\binom{|\C|-1}{\boldsymbol{\sigma}}\left(\frac{K}{L}\right)^{\ssum(\boldsymbol{\sigma})} \left(1-\frac{mK}{L}\right)^{|\C|-1-\ssum(\boldsymbol{\sigma})}
\end{equation}
otherwise.
\end{lemma}
\begin{proof}
%Observe that 
According to Lemma~\ref{lem:gamma-UI}(i), for $\s_g\in\C\setminus\{\s_0\}$, the maximum cross-correlation $H_{\s_0,\s_g}=1$.
We have $\sum_{\s_g\in\C\setminus\{\s_0\}}H_{\s_0,\s_g} \le |\C|-1 < {\gamma}n$, whenever $n> (|\C|-1)/\gamma$.
Therefore, by the Pigeonhole Principle, there is at least one successful packet transmitted among $x_1(\tau),\ldots,x_n(\tau)$ under $\gamma$-MPR model, leading to $P_n=1$.

Consider $1\leq n\leq \lfloor(|\C|-1)/\gamma\rfloor$.
%the sequences in $\C$ are chosen independently and 
Since the cyclic shifts are chosen independently and uniformly at random, by the Inclusion-Exclusion Principle, $P_{n}$ can be expressed as
\begin{equation}\label{eq:P_n-proof}
P_{n} = \sum_{m=1}^{n} (-1)^{m-1} \binom{n}{m} P(M_m),
\end{equation}
where $P(M_m)$ denotes the joint probability that a fixed subset $M_m$ of $m$ elements ($1 \le m \le n$) out of $\{x_1(\tau),\ldots,x_n(\tau)\}$ all achieve successful transmissions simultaneously.

For all $m$ packets in $M_m$ to succeed, each packet can experience at most $\gamma-1$ collisions from interfering sequences in $\C \setminus \{\s_0\}$. 
The collision state across these $m$ packets can be characterized by an $m$-tuple $\boldsymbol{\sigma} = (\sigma_1, \sigma_2, \dots, \sigma_m) \in \{0,1,\ldots,\gamma-1\}^m$, where $\sigma_i$ represents the exact number of interfering sequences colliding at the $i$-th packet in $M_m$. 
In this way, $\ssum(\boldsymbol{\sigma}) = \sum_{i=1}^{m} \sigma_i$ denotes the total number of collisions in the interfering sequences.

Because $H_{\s_0,\s_g}=1$ for $\s_g\in\C\setminus\{\s_0\}$, no single interfering sequence can collide with $\s_0$ at two or more of these $m$ packets, implying that the $\ssum(\boldsymbol{\sigma})$ collisions must be caused by $\ssum(\boldsymbol{\sigma})$ distinct sequences.
To evaluate $P(M_m)$, we sum over all valid collision configurations $\boldsymbol{\sigma}$:
\begin{enumerate}
    \item Out of the remaining $|\C|-1$ sequences, we select $\text{sum}(\boldsymbol{\sigma})$ distinct sequences and assign $\sigma_i$ sequences to collide with the $i$-th packet of $M_m$ for each $1 \leq i \leq m$. 
    The number of ways to partition these sequences is given by the multinomial coefficient $\binom{|\C|-1}{\boldsymbol{\sigma}}$.
    
    \item Under uniform random cyclic shifts, each selected sequence collides with its designated packet with probability $K/L$. 
    Thus, the probability that these $\text{sum}(\boldsymbol{\sigma})$ sequences collide as assigned is $(K/L)^{\ssum(\boldsymbol{\sigma})}$, due to the independence of these sequences.
    
    \item The remaining $|\C|-1-\ssum(\boldsymbol{\sigma})$ sequences must \emph{not} collide with any of the $m$ packets in $M_m$. 
    Since $H_{\s_0,\s_g}=1$, a sequence collides with a given packet with probability $K/L$ and hits none of the $m$ packets with probability $1 - m(K/L)$.
\end{enumerate}
Combining these events yields the explicit formula for $P(M_m)$:
\begin{align*}
    \sum_{\boldsymbol{\sigma}\in\{0,\ldots,\gamma-1\}^m}\binom{|\C|-1}{\boldsymbol{\sigma}} \left(\frac{K}{L}\right)^{\ssum(\boldsymbol{\sigma})} \left(1-\frac{mK}{L}\right)^{|\C|-1-\ssum(\boldsymbol{\sigma})}.
\end{align*}
Substituting $P(M_m)$ back into~\eqref{eq:P_n-proof} completes the proof.
\end{proof}

%%%%%%%%%%%%%%%%%%%%%%%%%%%%%%
\subsection{Average Individual Delay of $\s_0$}\label{sec:delay-0-s0}

Let $\mathbb{P}(E)$ denote the probability that an event $E$ occurs.

\begin{theorem}\label{thm:delay_0}
The average individual delay of $\s_0$ is given by the expected value
\begin{align*}
    \mathbb{E}[X_{\gamma;\C,0}] = \frac{p_K-1}{2} + p_K \sum_{n=1}^{\lfloor(|\C|-1)/\gamma\rfloor}(1-P_n),
\end{align*}
where $P_n$ is given in \eqref{eq:P_n_0}.
\end{theorem}
\begin{proof}
Recall that $\I_{\s_0}=\{0,p_K,\ldots,(q-1)p_K\}$.
Fix a cyclic shift $\tau\in\Z_L$ of $\s_0$, and consider the shifted characteristic set ${\I}_{\s_0}+\tau = \{x_1(\tau),x_2(\tau),\ldots,x_q(\tau)\}$, where $0\leq x_1(\tau)<x_2(\tau)<\cdots<x_q(\tau) < L$.
Since these positions are equally spaced, we have
%{\color{red} we may assume without loss of generality that}
\begin{equation}\label{eq:delay_0-0}
x_1(\tau)=\tau\bmod p_K, \text{ and } \ x_{n+1}(\tau)-x_n(\tau)=p_K
\end{equation}
for \(1\le n\le q-1\).

The conditional expected value of $X_{\gamma;\C,0}$ under the cyclic shift $\tau$ can be computed by
\begin{align}
\mathbb{E}[X_{\gamma;\C,0}|\tau] &= \sum_{t=0}^{\infty}\mathbb{P}(X_{\gamma;\C,0} > t|\tau) \notag \\ 
&= \sum_{t=0}^{L-1}\mathbb{P}\left(X_{\gamma;\C,0} > t|\tau\right) = \sum_{t=0}^{L-1} \left(1- \mathbb{P}\left(X_{\gamma;\C,0} \leq t|\tau\right)\right), \label{eq:delay_0-1}
\end{align}
where the second equality holds (i.e., the infinite summation being replaced by a finite one) due to the $\gamma$-UI property.
Obviously, the probability $\mathbb{P}(X_{\gamma;\C,0} \leq t|\tau)$ is equal to $0$ when $0\leq t<x_1(\tau)$ and is equal to $1$ when $t\geq x_q(\tau)$.
Suppose $x_n(\tau)\leq t <x_{n+1}(\tau)$ for some $n\in\{1,2,\ldots,q-1\}$.
Then, the probability $\mathbb{P}(X_{\gamma;\C,0} \leq t|\tau)$ turns out to be $P_n$. 
Therefore,
\begin{equation}\label{eq:delay_0-2}
\mathbb{P}(X_{\gamma;\C,0} \leq t|\tau) = 
\begin{cases}
    0, & \text{if } 0 \leq t < x_1(\tau), \\
    P_n, & \text{if } x_n(\tau) \leq t < x_{n+1}(\tau), n=1,\ldots,q-1, \\
    1, & \text{if } x_q(\tau) \le t < L.
\end{cases}
\end{equation}

Substituting~\eqref{eq:delay_0-2} into \eqref{eq:delay_0-1} yields
\begin{align*}
\mathbb{E}[X_{\gamma;\C,0}|\tau] &= x_1(\tau) + \sum_{n=1}^{q-1} (1-P_n)\left(x_{n+1}(\tau)-x_n(\tau)\right) \\
&= x_1(\tau) + p_K\sum_{n=1}^{q-1} (1-P_n).
\end{align*}
By Lemma~\ref{lem:P_n_0}, $P_n$ is independent of $\tau$ and equals $1$ whenever $n>(|\C|-1)/\gamma$.
Thus,
\begin{align*}
\mathbb{E}[X_{\gamma;\C,0}|\tau] = x_1(\tau) + p_K\sum_{n=1}^{\lfloor(|\C|-1)/\gamma\rfloor} (1-P_n).
\end{align*}

Finally, we have
\begin{align*}
\mathbb{E}[X_{\gamma;\C,0}] = \mathbb{E}[x_1(\tau)] + p_K\sum_{n=1}^{\lfloor(|\C|-1)/\gamma\rfloor} (1-P_n).
\end{align*}
%As, by~\eqref{eq:delay_0-0}, $\tau$ ranges uniformly over $\Z_L$, each value in $\{0,\ldots,p_K-1\}$ occurs equally often as $x_1(\tau)$. 
By~\eqref{eq:delay_0-0}, when $\tau$ is uniform over $\Z_L$, the value $x_1(\tau)$ is uniform over $\{0,\ldots,p_K-1\}$.
Therefore,
\begin{align*}
\mathbb{E}[x_1(\tau)] = \frac{p_K-1}{2},
\end{align*}
which gives the claimed formula. 
\end{proof}

%%%%%%%%%%%%%%%%%%%%%%%%%%%%%%
\subsection{Extension to CAC-Based Sequences}\label{sec:delay-0-CAC}

To extend the delay analysis to sequences with nonuniformly spaced transmission opportunities, we introduce the following notation. 
Let $\I=\{a_1,\ldots,a_w\}$ be any nonempty subset of $\Z_L$, where $0\le a_1<a_2<\cdots<a_w<L$.
Let $\boldsymbol{\delta}_{\I}=(\delta_1,\delta_2,\ldots,\delta_w)$ be the vector of consecutive differences of $\I$ by defining $\delta_1=L+a_1-a_w$, and $\delta_r=a_r-a_{r-1}$, for $2\leq r\leq w$.
Thus, $\delta_r$ is the cyclic gap immediately preceding $a_r$, and $\sum_{r=1}^{w}\delta_r=L$.

For \(d\in\{0,\ldots,w-1\}\), define
\begin{align*}
C_{\boldsymbol{\delta}_{\I}}(d)=\sum_{j=1}^{w}\delta_j\delta_{j\ominus d},
\end{align*}
where the subtraction $j\ominus d$ in the subscript of $\delta$ is the \emph{1-based modular subtraction} over $\{1,2,\ldots,w\}$, namely, the image of $j\ominus d$ is $\{1,2,\ldots,w\}$ rather than the standard $\{0,1,\ldots,w-1\}$.
More precisely, 
\begin{align*}
r\ominus d=1+((r-1-d)\bmod w).
\end{align*}
By reindexing the sum, we also have
\begin{align*}
C_{\boldsymbol{\delta}_{\I}}(d)=\sum_{j=1}^{w}\delta_j\delta_{j\oplus d},
\end{align*}
where the 1-based modular addition is given by 
\begin{align*}
r\oplus d=1+((r-1+d)\bmod w).
\end{align*}
These definitions apply to the characteristic set of any nonzero binary sequence of length \(L\).
The consecutive-difference vector and its cyclic autocorrelation were introduced in~\cite{WSLWS13,WSWS14}.
%Note that the notion $\boldsymbol{\delta}_{\I}$ and the concept $C_{\boldsymbol{\delta}_{\I}}(d)$ was first introduced in~\cite{WSLWS13}.

Returning to Example~\ref{ex:gamma=3}, we have $\boldsymbol{\delta}_{\I_{\s_4}}=(5,8,8,8,13,8,8,8)$, and $C_{\boldsymbol{\delta}_{\I_{\s_4}}}(0) = 578$, $C_{\boldsymbol{\delta}_{\I_{\s_4}}}(1) = 544$, and so on.

The preceding analysis also applies to $\gamma$-UI sequence sets obtained from CACs.
%The methodology of the derivation of $P_n$ and $\mathbb{E}[X_{\gamma;\C,0}]$ is applicable for evaluating the average individual delay of sequences in a $\gamma$-UI sequence set based on CACs.

\begin{corollary}\label{cor:delay_CAC}
Let $\C=\{\s_1,\ldots,\s_N\}$ be a $\gamma$-UI sequence set based on a CAC of weight $w$ and length $L$.
The average individual delay of $\s_g$ among $\C$ is given by the expected value
\begin{align*}
\mathbb{E}[X^\text{CAC}_{\gamma;\C,g}] = \frac{1}{L}\left(\frac{1}{2}C_{\boldsymbol{\delta}_{\I_{\s_g}}}(0) + \sum_{n=1}^{w-1}(1-P^*_n)C_{\boldsymbol{\delta}_{\I_{\s_g}}}(n) \right) - \frac{1}{2},
\end{align*}
where $P^*_n$ is given by
\begin{align*}
P^*_n = \sum_{m=1}^{n}(-1)^{m-1}\binom{n}{m} \sum_{\substack{\boldsymbol{\sigma}\in\{0,\ldots,\gamma-1\}^m\\ \ssum(\boldsymbol{\sigma})\leq N-1}}\binom{N-1}{\boldsymbol{\sigma}} \left(\frac{w}{L}\right)^{\ssum(\boldsymbol{\sigma})} \left(1-\frac{mw}{L}\right)^{N-1-\ssum(\boldsymbol{\sigma})}.
\end{align*}
\end{corollary}
\begin{proof}
We first claim that $P^*_n$ is the probability that the user equipped with $\s_g$ transmits at least one successful packet at the first $n$ transmission opportunities.
%The proof is similar to the one of Lemma~\ref{lem:P_n_0}.
Fix the cyclic shift $\tau\in\Z_L$ of the tagged sequence $\s_g$.
Consider any $m$, $1\leq m\leq n$, packets out of the first $n$ transmission opportunities.
Since, by the definition of CACs, $H_{\s_g,\s_h}=1$ for any $g\neq h$, each interfering sequence can overlap at most one of any $m$ specified packets of $\s_g$.
As every interfering sequence has weight $w$, it collides with each specified packet with probability $w/L$ and avoids all $m$ packets with probability $(1-mw/L)$. 
Applying the arguments in the proof of Lemma~\ref{lem:P_n_0} yields the stated formula for $P^*_n$.
In particular, $P^*_n$ depends only on $n$, not on the selected transmission positions or the cyclic shift $\tau$.

Write $\I_{\s_g}=\{a_1,\ldots,a_w\}$, with $0 \leq a_1<\cdots<a_w < L$, and let $\boldsymbol{\delta}_{\I_{\s_g}}=(\delta_1,\ldots,\delta_w)$. 
For a fixed shift $\tau$, write $\I_{\s_g}+\tau=\{x_1(\tau),\ldots,x_w(\tau)\}$, where $0 \leq x_1(\tau)<\cdots<x_w(\tau) < L$. 
Since \(\mathcal C\) has the \(\gamma\)-UI property, by the same argument in the proof of Theorem~\ref{thm:delay_0},
\begin{equation}\label{eq:delay_CAC_1}
\mathbb{E}[X^{\mathrm{CAC}}_{\gamma;\mathcal C,g}\mid\tau] = x_1(\tau) + \sum_{n=1}^{w-1}(1-P_n^*)\left(x_{n+1}(\tau)-x_n(\tau)\right).
\end{equation}
As $\tau$ ranges over $\Z_L$, the packet in $a_r$ becomes the first transmission opportunity for exactly $\delta_r$ shifts. 
For these shifts, $x_1(\tau)$ takes each value in $\{0,\ldots,\delta_r-1\}$ exactly once. 
Consequently,
\begin{equation}\label{eq:delay_CAC_2}
\sum_{\tau=0}^{L-1}x_1(\tau) = \sum_{r=1}^{w}\frac{\delta_r(\delta_r-1)}{2} = \frac{1}{2}\left(C_{\boldsymbol\delta_{\I_{\s_g}}}(0)-L\right).
\end{equation}
Moreover, whenever the packet in $a_r$ is the first transmission opportunity, the gap between the $n$-th and $(n+1)$-th opportunities is $\delta_{r\oplus n}$. 
Hence,
\begin{equation}\label{eq:delay_CAC_3}
\sum_{\tau=0}^{L-1} \bigl(x_{n+1}(\tau)-x_n(\tau)\bigr) = \sum_{r=1}^{w}\delta_r\delta_{r\oplus n} = C_{\boldsymbol\delta_{\I_{\s_g}}}(n).
\end{equation}
Averaging the conditional expectation in~\eqref{eq:delay_CAC_1} over all $L$ shifts and substituting the two identities in \eqref{eq:delay_CAC_2} and \eqref{eq:delay_CAC_3} proves the result. 
\end{proof}

%%%%%%%%%%%%%%%%%%%%%%%%%%%%%%
\section{Average Individual Delay of the Remaining $\gamma$-CRT Sequences}\label{sec:delay-g}

We now consider the remaining sequences $\s_g\in\C$, where $1\leq g\leq p_K$, under the delay model introduced in Section~\ref{sec:delay-0}. 
Unlike the case of $\s_0$,, the probability of success within the first $n$ transmission opportunities may depend on the tagged-sequence shift. 
To account for this dependence, we first count collision configurations, then compute joint success probabilities, and finally average the resulting waiting times over all cyclic shifts.

%%%%%%%%%%%%%%%%%%%%%%%%%%%%%%
\subsection{Collision-Counting Quantities}\label{sec:delay-g-quantity}

%We next analyze the individual delay of a tagged sequence $\s_g\in\C$, where $1\leq g\leq p_K$.
Fix its cyclic shift $\tau\in\mathbb{Z}_L$, and write $\I_{\s_g}+\tau=\{x_1(\tau),x_2(\tau),\ldots,x_K(\tau)\}$, where $0\leq x_1(\tau)<x_2(\tau)<\cdots<x_K(\tau)<L$.
For $1\leq n\leq K$, let $P_{n,g}(\tau)$ denote the probability that at least one of the first $n$ transmission opportunities is successful under the $\gamma$-MPR model, with probability taken over the independent uniform shifts of the interfering sequences. 
Unlike the case of $\s_0$, this probability may depend on the tagged-sequence shift $\tau$.

To characterize $P_{n,g}(\tau)$, fix the tagged-sequence shift $\tau\in\Z_L$ and an interfering sequence $\s_h\in\C\setminus\{\s_g\}$. 
For $B,D\subseteq\I_{\s_g}+\tau$, define two collision-counting quantities by
\begin{equation*}%\label{eq:alpha_gh}
\alpha^{(\tau)}_{g,h}(B)\triangleq\left|\left\{\tau_h\in\mathbb{Z}_L:\, (\I_{\s_g}+\tau)\cap(\I_{\s_h}+\tau_h)=B\right\}\right|
\end{equation*}
and
\begin{equation*}%\label{eq:beta_gh}
\beta^{(\tau)}_{g,h}(D)\triangleq\left|\left\{\tau_h\in\mathbb{Z}_L:D\subseteq
(\I_{\s_g}+\tau)\cap(\I_{\s_h}+\tau_h)\right\}\right|.
\end{equation*}
Thus, $\alpha^{(\tau)}_{g,h}(B)$ counts the interfering shifts whose exact collision set is $B$, while $\beta^{(\tau)}_{g,h}(D)$ counts those producing collisions at every position in $D$, allowing additional collisions outside $D$.

Both quantities are invariant under a common translation of the tagged characteristic set and the specified subset, since the map $\tau_h\mapsto\tau_h+\eta$ gives a bijection between the corresponding interfering shifts. 
In other words, for any $\eta\in\Z_L$,
\begin{align*}
\alpha_{g,h}^{(\tau+\eta)}(B+\eta) = \alpha_{g,h}^{(\tau)}(B) \quad\text{and}\quad \beta_{g,h}^{(\tau+\eta)}(D+\eta) = \beta_{g,h}^{(\tau)}(D),
\end{align*}
where all additions are taken modulo $L$. 
Whenever $\tau$ is fixed in the discussion, we suppress the superscript and write $\alpha_{g,h}(B)$ and $\beta_{g,h}(D)$ without ambiguity.

\begin{comment}
To characterize $P_{n,g}(\tau)$, we first introduce the following exact collision-counting quantity. 
For an interfering sequence $\s_h\in C\setminus\{\s_g\}$, and $B\subseteq \I_{\s_g}+\tau$, define
\begin{equation}\label{eq:alpha_gh}
\alpha_{g,h}(B)\triangleq\left|\left\{\tau_h\in\mathbb{Z}_L:\, (\I_{\s_g}+\tau)\cap(\I_{\s_h}+\tau_h)=B\right\}\right|.
\end{equation}
Thus, $\alpha_{g,h}(B)$ counts the shifts of $\s_h$ for which the exact collision set with the tagged sequence is $B$.
The tagged-sequence shift $\tau$ is fixed and suppressed in the notation.
%Note that this quantity does not depend on the shift $\tau$.

For comparison, define the number of shifts that produce collisions at every position in a specified subset $D\subseteq\I_{\s_g}+\tau$, while allowing additional collisions outside $D$, by
\begin{equation*}%\label{eq:beta_gh}
\beta_{g,h}(D)\triangleq\left|\left\{\tau_h\in\mathbb{Z}_L:D\subseteq
(\I_{\s_g}+\tau)\cap(\I_{\s_h}+\tau_h)\right\}\right|.
\end{equation*}
Both counting quantities are invariant under a common translation of the tagged characteristic set and the specified subset, since translating the interfering shift by the same amount gives a bijection between the corresponding shifts.
%This quantity does not depend on the shift $\tau$ either.
\end{comment}

\begin{lemma}
Fix $\s_g\in \C$ with $1\leq g\leq p_K$ and a cyclic shift $\tau\in\mathbb{Z}_L$. 
For any $B\subseteq \I_{\s_g}+\tau$ and $\s_h\in \C\setminus\{\s_g\}$, one has
\begin{equation}\label{eq:alpha_from_beta}
\alpha_{g,h}(B)=\sum_{\substack{\,B\subseteq D\subseteq \I_{\s_g}+\tau}}(-1)^{|D|-|B|}\beta_{g,h}(D).
\end{equation}
\end{lemma}
\begin{proof}
For a fixed subset $D\subseteq \I_{\s_g}+\tau$, consider a cyclic shift $\tau_h$ of the interfering sequence $\s_h$ such that $D\subseteq(\I_{\s_g}+\tau)\cap(\I_{\s_h}+\tau_h)$, namely, $\tau_h$ contributes $1$ to the quantity $\beta_{g,h}(D)$.
For this particular shift, let
\[B=(\I_{\s_g}+\tau)\cap(\I_{\s_h}+\tau_h)\]
denote the exact collision set. 
Clearly, $D\subseteq B$.

Conversely, every cyclic shift $\tau_h$ of $\s_h$ whose exact collision set is some $B\supseteq D$ is counted once in $\beta_{g,h}(D)$. 
Since the sets of shifts corresponding to different exact collision sets $B$ are disjoint, summing over all $B\supseteq D$ gives
\begin{equation*}%\label{eq:beta_alpha_relation}
\beta_{g,h}(D)=\sum_{\substack{\,D\subseteq B\subseteq \I_{\s_g}+\tau}}\alpha_{g,h}(B).
\end{equation*}
Equation \eqref{eq:alpha_from_beta} then follows from the M\"obius Inversion Formula on the subset lattice, whose M\"obius function is
\[\mu(B,D)=(-1)^{|D|-|B|}.\] This completes the proof.
\end{proof}

%%%%%%%%%%%%%%%%%%%%%%%%%%%%%%
\subsection{Joint Success Probabilities}\label{sec:delay-g-Png}

For each $\s_h\in \C\setminus\{\s_g\}$, let $B_h\subseteq \I_{\s_g}+\tau$ denote its exact collision set with the tagged sequence $\s_g$.
That is, $B_h = (\I_{\s_g}+\tau)\cap(\I_{\s_h}+\tau_h)$ contains all packet positions of $\s_g$ at which $\s_h$ collides with $\s_g$ under a given cyclic shift pair $(\tau,\tau_h)$.
For $x\in \I_{\s_g}+\tau$, define the indicator function
\begin{equation*}
\mathbf{1}_{\{x\in B_h\}}\triangleq
\begin{cases}
1, &x\in B_h,\\
0, &x\notin B_h.
\end{cases}
\end{equation*}
Note that $\mathbf{1}_{\{x\in B_h\}}$ indicates whether the interfering sequence $\s_h$ collides with the packet of $\s_g$ transmitted at $x$.
For a nonempty subset $A\subseteq \I_{\s_g}+\tau$, define
\begin{equation}\label{eq:B_gamma}
\begin{aligned}
\mathcal{B}_{\gamma,g}(A)\triangleq\Bigg\{&(B_h)_{\s_h\in \C\setminus\{\s_g\}}:\sum_{\s_h\in \C\setminus\{\s_g\}}\mathbf{1}_{\{x\in B_h\}}\leq \gamma-1,\quad \forall x\in A\Bigg\}.
\end{aligned}
\end{equation}
An element $(B_h)_{\s_h\in \C\setminus\{\s_g\}}$ of $\mathcal{B}_{\gamma,g}(A)$ represents a collision configuration of all interfering sequences. 
For each $x\in A$, the quantity $\sum_{\s_h\in \C\setminus\{\s_g\}}\mathbf{1}_{\{x\in B_h\}}$ counts the number of interfering sequences that collide with the packet transmitted by $\s_g$ at $x$. 
Hence, the condition in \eqref{eq:B_gamma} ensures that every packet in $A$ experiences at most $\gamma-1$ interfering transmissions and is therefore successfully received under the $\gamma$-MPR model.

Using the above collision configurations, we next compute the probability that a given subset of packets is successfully
received.

\begin{lemma}
Fix $\s_g\in \C$ with $1\leq g\leq p_K$ and a cyclic shift $\tau\in\mathbb{Z}_L$. 
For a nonempty subset $A\subseteq \I_{\s_g}+\tau$, let $E_x$ denote the event that the packet transmitted by $\s_g$ at $x \in A$ is successfully received. 
Then
\begin{equation}\label{eq:joint_success_sg}
\mathbb{P}\left(\bigcap_{x\in A}E_x\right)=\frac{1}{L^{|\C|-1}}\sum_{\mathbf B\in\mathcal B_{\gamma,g}(A)}\prod_{\s_h\in \C\setminus\{\s_g\}}\alpha_{g,h}(B_h),
\end{equation}
where $\mathcal B_{\gamma,g}(A)$ is given in \eqref{eq:B_gamma}.
Consequently, writing $x_i=x_i(\tau)$ for brevity, the probability $P_{n,g}(\tau)$ that at least one of the first $n$ transmitted packets is successfully received is
\begin{equation}\label{eq:Png_general}
P_{n,g}(\tau)=\sum_{\emptyset\neq A\subseteq\{x_1,\ldots,x_n\}}(-1)^{|A|+1}\frac{1}{L^{|\C|-1}}\sum_{\mathbf B\in\mathcal B_{\gamma,g}(A)}\prod_{\s_h\in \C\setminus\{\s_g\}}\alpha_{g,h}(B_h).
\end{equation}
\end{lemma}
\begin{proof}
For each interfering sequence $\s_h\in \C\setminus\{\s_g\}$ and each subset $B_h\subseteq \I_{\s_g}+\tau$, by definition, there are $\alpha_{g,h}(B_h)$ relative cyclic shifts of $\s_h$ that produce the exact collision set $B_h$.
Since the relative cyclic shifts of the interfering sequences are independent and uniformly distributed over $\mathbb{Z}_L$, a collision configuration $\mathbf B=(B_h)_{\s_h\in \C\setminus\{\s_g\}}$ occurs with probability
\begin{equation}\label{eq:collision_config_probability}
\mathbb{P}(\mathbf B)=\frac{1}{L^{|\C|-1}}\prod_{\s_h\in C\setminus\{\s_g\}}\alpha_{g,h}(B_h).
\end{equation}

Consider a packet position $x\in A$. Under the collision configuration $\mathbf B$, the number of interfering sequences transmitting simultaneously with $\s_g$ at $x$ is
\begin{equation*}\label{eq:number_interferers_delay}
N_g(x;\mathbf B)=\sum_{\s_h\in \C\setminus\{\s_g\}}\mathbf{1}_{\{x\in B_h\}}.
\end{equation*}
Under the $\gamma$-MPR model, all packets in $A$ are successfully received if and only if \[N_g(x;\mathbf B)\leq\gamma-1,\qquad\forall x\in A,\]
which is precisely the condition $\mathbf B\in\mathcal B_{\gamma,g}(A)$.

Summing \eqref{eq:collision_config_probability} over all admissible collision configurations gives
\begin{equation*}
\mathbb{P}\left(\bigcap_{x\in A}E_x\right)=\frac{1}{L^{|\C|-1}}\sum_{\mathbf B\in\mathcal B_{\gamma,g}(A)}\prod_{\s_h\in \C\setminus\{\s_g\}}\alpha_{g,h}(B_h),
\end{equation*}
which proves \eqref{eq:joint_success_sg}.

Finally, applying the Inclusion--Exclusion Principle to the events $E_{x_1},E_{x_2},\ldots,E_{x_n}$ yields
\begin{equation}\label{eq:P_{n,g}}
P_{n,g}(\tau)=\sum_{\emptyset\neq A\subseteq\{x_1,\ldots,x_n\}}(-1)^{|A|+1}\mathbb{P}\left(\bigcap_{x\in A}E_x\right).
\end{equation}
Substituting \eqref{eq:joint_success_sg} into \eqref{eq:P_{n,g}} completes the proof.
\end{proof}

%%%%%%%%%%%%%%%%%%%%%%%%%%%%%%
\subsection{Average Individual Delay}\label{sec:delay-g-sg}

Write $\I_{\s_g}=\{a_1,a_2,\ldots,a_K\}$, where $0=a_1<\cdots<a_K<L$, and let $\boldsymbol{\delta}_{\I_{\s_g}}=(\delta_1,\ldots,\delta_K)$ be its consecutive-difference vector.

For $1\leq r\leq K$, define
\begin{align*}
\mathcal{T}_r \triangleq \{(k-a_r)\bmod L:k=0,\ldots,\delta_r-1\}.
\end{align*}
The sets $\mathcal{T}_1,\ldots,\mathcal{T}_K$ partition $\Z_L$. 
For every $\tau\in\mathcal{T}_r$, the transmission position originating from $a_r$ is the first transmission opportunity, and $|\mathcal{T}_r|=\delta_r$.

Within each $\mathcal{T}_r$, the first $n$ transmission positions differ only by a common translation. 
Since the interfering shifts are independent and uniform, this translation does not change their joint success probabilities. 
Thus, $P_{n,g}(\tau)$ is constant over $\mathcal{T}_r$, and we define
\begin{equation}\label{eq:P_{n,g}^r}
P_{n,g}^{(r)}\triangleq P_{n,g}(\tau),\qquad \tau\in\mathcal{T}_r.
\end{equation}

\begin{theorem}\label{thm:delay_g}
For $\s_g\in\C$ with $1\leq g\leq p_K$, the average individual delay of $\s_g$ is
\begin{equation}\label{eq:average_delay_sg}
\mathbb{E}[X_{\gamma;\C,g}]=\frac{1}{L}\left(\frac{1}{2}C_{\boldsymbol{\delta}_{\I_{\s_g}}}(0)+\sum_{n=1}^{K-1}\sum_{r=1}^{K}\left(1-P_{n,g}^{(r)}\right)\delta_r\delta_{r\oplus n}\right)-\frac{1}{2},
\end{equation}
where $P_{n,g}^{(r)}$ is given in~\eqref{eq:Png_general} and \eqref{eq:P_{n,g}^r} and $\oplus$ denotes the $1$-based modular addition over $\{1,2,\ldots,K\}$.
\end{theorem}
\begin{proof}
Fix the cyclic shift $\tau\in\mathbb{Z}_L$ of $\s_g$.
By the same waiting-time decomposition used in the proof of Theorem~\ref{thm:delay_0}, and since $\C$ has the $\gamma$-UI property,
\begin{equation}\label{eq:delay_g-1}
\mathbb{E}[X_{\gamma;\C,g}\mid\tau]=x_1(\tau)+\sum_{n=1}^{K-1}\left(1-P_{n,g}(\tau)\right)\left(x_{n+1}(\tau)-x_n(\tau)\right).
\end{equation}

The cyclic-gap identity in~\eqref{eq:delay_CAC_2} applies to
any nonempty characteristic set.
Hence, 
\begin{equation}\label{eq:delay_g-2}
\sum_{\tau=0}^{L-1}x_1(\tau)=\frac{1}{2}\left(C_{\boldsymbol{\delta}_{\I_{\s_g}}}(0)-L\right).
\end{equation}

For every $\tau\in\mathcal{T}_r$, the first transmission opportunity originates from $a_r$.
Consequently, $x_{n+1}(\tau)-x_n(\tau)=\delta_{r\oplus n}$ and $P_{n,g}(\tau)=P_{n,g}^{(r)}$.
Since $|\mathcal{T}_r|=\delta_r$, grouping the shifts by $\mathcal{T}_r$ gives
\begin{equation}\label{eq:delay_g-3}
\sum_{\tau=0}^{L-1}\left(1-P_{n,g}(\tau)\right)\left(x_{n+1}(\tau)-x_n(\tau)\right)=\sum_{r=1}^{K}\left(1-P_{n,g}^{(r)}\right)\delta_r\delta_{r\oplus n}.
\end{equation}
This is the weighted counterpart of the cyclic-gap identity in~\eqref{eq:delay_CAC_3}.

Averaging the conditional expectation in~\eqref{eq:delay_g-1} over all $L$ shifts and substituting the identities \eqref{eq:delay_g-2} and \eqref{eq:delay_g-3} yields the stated formula.
\end{proof}

When $K$ is prime, the distinguished sequence $\s_{p_K}=\s_K$ has equally spaced transmission opportunities, leading to the following simplification.

\begin{corollary}
Assume that $\gamma\geq 3$, $K$ is prime, and $\s_K\in\C$. 
Then the average individual delay of $\s_{K}$ is given by
\begin{align*}
\mathbb{E}[X_{\gamma;\C,K}]=\frac{q-1}{2}+ q\sum_{n=1}^{K-1}(1-P_{n,K}). 
\end{align*}
Here $P_{n,K}$ denotes the common value of $P_{n,K}(\tau)$, which is independent of $\tau$ and given in~\eqref{eq:Png_general}. %can be evaluated using \eqref{eq:P_{n,g}}.
\end{corollary}
\begin{proof}
In this case, we have $L=Kq$ and $\I_{\s_K}=\{0,q,2q,\ldots,(K-1)q\}$, where $q=\lceil 2K/\gamma \rceil < K$. 
Hence, all elements in $\boldsymbol{\delta}_{\I_{\s_K}}$ equal $q$, and
\begin{align*}
C_{\delta_{I_{s_K}}}(d)=Kq^2
\end{align*}
for every $d=0,1,\ldots,K-1$.

For any cyclic shift $\tau$ of $\s_K$, the first $n$ transmission positions are a common translation of $\{0,q,\ldots,(n-1)q\}$. 
The uniform interfering shifts therefore imply that $P_{n,K}(\tau)$ is independent of $\tau$. 
Denote this common value by $P_{n,K}$.

Applying Theorem~\ref{thm:delay_g} gives
\begin{align*}
\mathbb{E}[X_{\gamma;C,K}]
&=
\frac{1}{Kq}\left(\frac{Kq^2}{2} + Kq^2\sum_{n=1}^{K-1}(1-P_{n,K})\right)-\frac{1}{2} \\
&=
\frac{q-1}{2}+ q\sum_{n=1}^{K-1}(1-P_{n,K}),
\end{align*}
as desired. 
\end{proof}

%%%%%%%%%%%%%%%%%%%%%%%%%%%%%%
\section{Concluding Remarks}\label{sec:conclusion}

This paper developed a unified CRT-based construction, $\gamma$-CRT$(K)$, for $\gamma$-UI sequences. 
We established pairwise cross-correlation bounds and proved that every subset of at most $K$ constructed sequences has the $\gamma$-UI property. 
This guarantee also holds for every subset of $k+1$ sequences containing at least one of the two distinguished sequences, $\s_0$ and $\s_{p_K}$. 
Under independent uniform cyclic shifts, we derived exact expressions for the average individual delay by exploiting the equal spacing of the transmission slots of $\s_0$ and developing a collision-counting method for the remaining sequences. 
We also extended the delay analysis to CAC-based $\gamma$-UI sequences and established additional results for new prime and classical prime sequences.

An important direction for future research is to determine $L(N,\gamma)$, the minimum common length of a set of $N$ sequences with the $\gamma$-UI property.  
For $\gamma=1$, the known lower bound and the classical CRT construction give $8N^2/9\leq L(N,1)\leq p_N(2N-1)$ \cite{SZW10,SWSC09}. 
For fixed $\gamma\geq 2$, the CAC-based result in~\cite[Theorem 12]{ZLSW14} establishes $2N^2/\gamma$ as a liminf benchmark under the additional restrictions $N$ is divisible by $\gamma$ and that every sequence has weight $N/\gamma$.
Our extended user-support guarantee allows us to take $K=N-1$ and select $N$ sequences containing at least one of the two distinguished sequences. 
Together with the new prime sequences for the case where $\gamma=2$ and $N-1$ is prime, this yields
\begin{align*}
L(N,\gamma)\leq p_{N-1}\left\lceil\frac{2(N-1)}{\gamma}\right\rceil = \left(\frac{2}{\gamma}+o(1)\right)N^2,
\end{align*}
where the inequality holds for $N\geq\gamma+1$, and the asymptotic expression is for fixed $\gamma$ as $N\to\infty$. 
Establishing lower bounds without weight restrictions and determining whether this leading constant can be improved remain important open problems. 
Constructions with shorter lengths for finite $N$ would also be valuable.

%This paper has developed a unified CRT-based construction, $\gamma$-CRT$(K)$, for $\gamma$-UI sequences. 
%We proved that every subset of at most $K$ constructed sequences has the $\gamma$-UI property, and that this guarantee extends to subsets of $K+1$ sequences containing at least one of the two distinguished sequences. 
%Under independent uniform cyclic shifts, we derived exact expressions for the average individual delay, using equal spacing for $\s_0$ and collision counting for the remaining sequences. 
%We also extended the delay analysis to $\gamma$-UI sequence sets obtained from constant-weight CACs.

%%%%%%%%%%%%%%%%%%%%%%%%%%%%%%
\appendices 

%%%%%%%%%%%%%%%%%%%%%%%%%%%%%%
\section{New Prime Sequences and their $2$-UI Property}\label{app:2UI_proof}

We first recall the \emph{new prime sequences} in~\cite{SXWLZM26}.

\begin{definition}[New Prime Sequences~\cite{SXWLZM26}]\label{def:NPS} \rm
Let $p$ be a prime number. 
The set of new prime sequences, denoted by $\mathcal{S}_{\NP}(p)=\{\s^{\NP}_0,\s^{\NP}_1,\ldots,\s^{\NP}_p\}$, consists of $p+1$ binary sequences of length $L = p^2$, defined via their characteristic sets as follows.
For $g=0,1,...,p-1$, let 
\begin{equation}\label{eq:NPS-def-g}
\I_{\s^{\text{NP}}_g} \triangleq \{j(1+gp)\bmod p^2:\, j=0,1,...,p-1 \};    
\end{equation}
and let 
\begin{equation}\label{eq:NPS-def-p}
\I_{\s^{\NP}_p} \triangleq \{jp \bmod p^2:\, j=0,1,...,p-1 \}.    
\end{equation}
\end{definition}

It was shown in~\cite{SXWLZM26} that every subset of $p$ sequences chosen from $\mathcal{S}_{\text{NP}}(p)$ has the 2-UI property. 
We strengthen this result by showing that the \emph{entire} family of $p+1$ sequences has this property.
This extension implies that $\mathcal{S}_{\NP}(p)$ can support up to $p+1$ active users under the $2$-UI condition; in other words, potential users can all be active users in this case.

The Hamming cross-correlation between any two distinct sequences in $\mathcal{S}_{\NP}(p)$ is characterized in ~\cite[Lemma 1]{SXWLZM26}: 
\begin{align*}
    H_{\s^{\NP}_g,\s^{\NP}_h} = \begin{cases}
        1, & \text{if } g=p \text{ or } h=p, \\
        2, & \text{if } 0 \leq g,h \leq p-1.
    \end{cases}
\end{align*}
For $0\leq g\leq p-1$ (resp., $g=p$), the set $\I_{\s^{\NP}_g}$ has at most $2(p-1)+1=2p-1$ overlaps (resp., $p$ overlaps) with  $\cup_{h\neq g} (\I^{\NP}_{\s_h}+\tau_h)$ for any $\tau_h$, counting multiplicity.
Since $|\I_{\s^{\NP}_g}|=p$, the Pigeonhole Principle guarantees at least one element of $\I_{\s^{\NP}_g}$ that is covered by at most one of the other $p$ shifted characteristic sets. 
Thus, every sequence has a successful transmission under arbitrary relative shifts, proving the $2$-UI property of $\mathcal{S}_{\NP}(p)$.

%%%%%%%%%%%%%%%%%%%%%%%%%%%%%%
\section{Equivalence Between New Prime and Classical Prime Sequences}\label{app:isomorphism}

%The set $\{\s^{\NP}_1,\ldots,\s^{\NP}_p\}$ of $p$ sequences in $\mathcal{S}_{\NP}(p)$ is structurally isomorphic to the original \emph{prime sequences} introduced in~\cite{YK95}. 
%Since~\cite{SXWLZM26} proved the $2$-UI property for these $p$ sequences, this isomorphism implicitly yields a substantially simpler proof for the $2$-UI property of the original prime sequences. 

We show that the subset $\{\s^{\NP}_1,\ldots,\s^{\NP}_p\}$ of the new prime sequences defined in Definition~\ref{def:NPS} coincides with the classical prime sequences in~\cite{YK95} after reindexing. 
Together with the $2$-UI result in Appendix~\ref{app:2UI_proof}, this correspondence gives an alternative proof of the $2$-UI property of the classical prime sequences. 
We begin by recalling their definition.

\begin{definition}[Prime Sequences~\cite{YK95}]\label{def:PS} \rm
Let $p$ be a prime number. 
The set of prime sequences, denoted by $\mathcal{S}_{\PP}(p)=\{\s^{\PP}_0,\s^{\PP}_1,\ldots,\s^{\PP}_{p-1}\}$, consists of $p$ binary sequences of length $L = p^2$, defined via their characteristic sets as:
For $i=0,1,\ldots,p-1$, let
\begin{equation}\label{eq:PS-def}
\I_{\s^{\PP}_i} \triangleq \{jp + (ij\ \bmod {p}):\, j=0,1,...,p-1 \}.    
\end{equation}
\end{definition}

Define the mapping $\psi:\{\s^{\NP}_1, \ldots, \s^{\NP}_p\}\to\mathcal{S}_{\PP}(p)$ by
\begin{equation}
    \psi(\s^{\NP}_g) = \begin{cases}
        \s^{\PP}_0 & \text{if } g=p, \\
        \s^{\PP}_{g^{-1} \bmod{p}} & \text{if } 1\leq g\leq p-1,
    \end{cases}
\end{equation}
where $g^{-1}$ denotes the multiplicative inverse of $g$ in the finite field $\F_p$.
This mapping is a bijection.
We now show that each sequence and its image have identical characteristic sets.

It is easy to see from \eqref{eq:NPS-def-p} and \eqref{eq:PS-def} that
\begin{align*}
    \I_{\s^{\NP}_p} = \{0,p,2p,\ldots,(p-1)p\} = \I_{\s^{\PP}_0}.
\end{align*}
The case when $g=p$ is completed.

For $1 \leq g \leq p-1$, the characteristic set of $\s^{\NP}_g$ is $\I_{\s^{\NP}_g} = \{ k(1+gp) \bmod{p^2} :\, k=0,1,\ldots,p-1 \}$. 
To establish the exact correspondence, we assert that an arbitrary element in $\I_{\s^{\NP}_g}$ must perfectly match a corresponding element in $\I_{\s^{\PP}_i}=\{jp + (ij\ \bmod {p}):\, j=0,1,...,p-1 \}$, and show the index $i$ is equal to $g^{-1} \bmod{p}$.
By definitions, we have the following congruence:
\begin{equation}\label{eq:isomorphism-proof-1}
k(1+gp) \equiv [jp + (ij \bmod{p})] \quad \bmod{p^2},
\end{equation}
where $k, j \in \{0, 1, \ldots, p-1\}$. 
As each characteristic set contains $0$, we only consider $1\leq k,j\leq p-1$; in other words, we set the two $0$'s in $\I_{\s^{\NP}_g}$ and $I_{\s^{\PP}_i}$ matched.
Since $1\leq g\leq p-1$, by comparing the base-$p$ digits on both sides of~\eqref{eq:isomorphism-proof-1} with the $p$-ary representations, we get
\begin{equation}\label{eq:isomorphism-proof-2}
    k\equiv ij \bmod{p} \quad \text{and} \quad kg\equiv j \bmod{p}.
\end{equation}
By substituting the first congruence into the second one, we obtain $ijg\equiv j \bmod{p}$, and thus $ig=1$ in $\F_p$ due to the invertibility of $j$.
Now, we get 
\begin{align*}
    i=g^{-1} \quad \text{in }\F_p.
\end{align*}
Furthermore, substituting this derived $i$ back into the first congruence in~\eqref{eq:isomorphism-proof-2} yields the exact index mapping between the two sets: $\{k:\,k=1,\ldots,p-1\}$ and $\{g^{-1}j:\,j=1,\ldots,p-1\}$. 
%This step-by-step derivation proves that two sets of sequences in $\s^{\NP}_g$ and $\s^{\PP}_{g^{-1}}$ are identical.
Therefore, $\I_{\s^{\NP}_g}=\I_{\s^{\PP}_{g^{-1}}}$ for every $1\leq g\leq p-1$.
Together with $\I_{\s^{\NP}_p}=\I_{\s^{\PP}_0}$, this proves the claimed equivalence.

Take $p=5$ as an example.
One can check that
\begin{align*}
\I_{\s^{\NP}_1} &= \{0,6,12,18,24\} = \I_{\s^{\PP}_1} \quad (\text{since } 1^{-1} \equiv 1 \text{ in }\F_5), \\
\I_{\s^{\NP}_2} &= \{0,11,22,8,19\} = \I_{\s^{\PP}_3} \quad (\text{since } 2^{-1} \equiv 3 \text{ in }\F_5), \\
\I_{\s^{\NP}_3} &= \{0,16,7,23,14\} = \I_{\s^{\PP}_2} \quad (\text{since } 3^{-1} \equiv 2 \text{ in }\F_5), \\
\I_{\s^{\NP}_4} &= \{0,21,17,13,9\} = \I_{\s^{\PP}_4} \quad (\text{since } 4^{-1} \equiv 4 \text{ in }\F_5), \\
\I_{\s^{\NP}_5} &= \{0,5,10,15,20\} = \I_{\s^{\PP}_0}.
\end{align*}

%%%%%%%%%%%%%%%%%%%%%%%%%%%%%%
\bibliography{Ref_UI-MPR}
\bibliographystyle{IEEEtran}

\end{document}